\documentclass[prb,aps,twocolumn,amsmath,amsfonts]{revtex4}   
\usepackage{epsfig,psfrag}
\usepackage{amsmath}
\usepackage{latexsym,exscale}
\usepackage{float}
\usepackage{graphics,subfigure}
\usepackage{color}
\usepackage{lscape}
\usepackage[ansinew]{inputenc}
\usepackage{pstricks, pst-tree, pst-node}
\newcommand{\ket}[1]{\left |#1\right \rangle}
\newcommand{\bra}[1]{\left \langle #1\right|}

\begin{document}
\title{Highest weight state description of the isotropic spin-$1$ chain}
\author{Markus Andres, Imke Schneider, and Sebastian Eggert} 
\affiliation{Department of Physics, University of Kaiserslautern,  
D-67663 Kaiserslautern, Germany}
\begin{abstract}
We introduce an overcomplete highest weight state basis as a  calculational tool
for the description of the isotropic spin-$1$ chain 
with bilinear exchange coupling $J_1$
and biquadratic coupling $J_2$.
The ground state can be expressed exactly
at the three special points in the phase diagram
where the  Hamiltonian corresponds to
a sum of nearest neighbor total spin projection operators
($J_1=0>J_2$, $J_1=-J_2<0$, and $J_1=-J_2/3>0$).
In particular, at the phase transition point $J_1=-J_2<0$ 
it is possible to exactly compute the ground states, 
excited states, expectation values, and
correlation functions by using the new total spin basis.
\end{abstract}

\maketitle
\section{Introduction}
\label{intro}
\begin{figure}
\begin{center}
\includegraphics[width=0.45\textwidth]{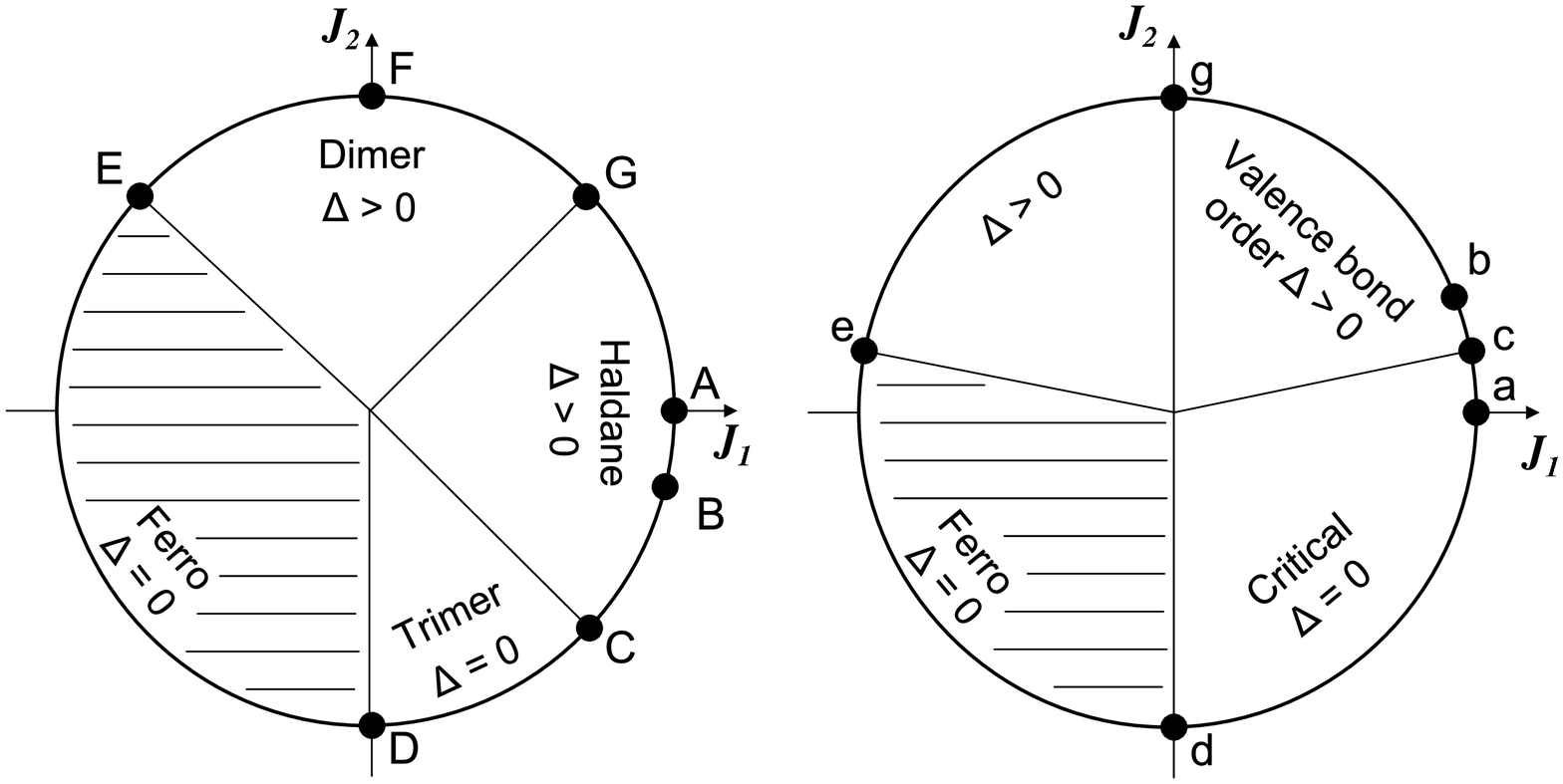}
\end{center}
\caption{\textit{{\bf lhs}: Phase diagram of the spin-$1$ chain}: 
$A$: Pure Heisenberg chain; 
$B$: AKLT point [\onlinecite{AKLT1,AKLT2}]; 
$C$: Sutherland model [\onlinecite{Sutherland}]; 
$D$: Phase transition [\onlinecite{Parkinson}]; 
$E$: Phase transition [\onlinecite{FathSolyom3,Chubukov}]; 
$F$: Exactly solvable [\onlinecite{Klumper1,Klumper2,Klumper3}]; 
$G$: $SU(2)_2$ integrable point [\onlinecite{Takhtajan,Babujian1,Babujian2}].
\textit{{\bf rhs}: Phase diagram of the spin-$\frac{1}{2}$ chain}: 
$a$: Heisenberg chain [\onlinecite{Mattis}]; 
$b$: Majumdar-Ghosh point [\onlinecite{MajumdarGhosh}]; 
$c$: Critical point [\onlinecite{Eggert}]; 
$d$: Phase transition; 
$g$: Two independent chains, $SU(2)_1\times SU(2)_1$ integrable; 
$e$: Phase transition [\onlinecite{Natsume}]. 
The points $B$, $D$, and $E$ are ''projection points'' treated in this paper.}
\label{phasediagram}
\end{figure}

There has been a large interest in the isotropic one-dimensional
spin-$1$ chain ever since
Haldane's prediction,\cite{Haldane1,Haldane2} that the excitation spectrum in integer
spin Heisenberg chains should show a gap in strong contrast to the model with half
integer spins.  The general $SU(2)$ invariant spin-1 chain model with nearest
neighbor coupling is given by
\begin{eqnarray}\label{model}
H&=&\sum_{i=1}^N \left(J_1 \textbf{S}_i\cdot
\textbf{S}_{i+1}-J_2(\textbf{S}_i\cdot \textbf{S}_{i+1})^2\right)
\nonumber \\
&=&J\sum_{i=1}^N \left(\cos \Theta \textbf{S}_i\cdot \textbf{S}_{i+1}-\sin
\Theta (\textbf{S}_i\cdot \textbf{S}_{i+1})^2\right)\ ,
\end{eqnarray}
where $\textbf{S}_i$ are the spin-1
operators at the site $i$ in a one-dimensional periodic lattice with $N$ sites.
Exact
analytical solutions at special points were obtained by Affleck, Kennedy, Lieb
and Tasaki (AKLT),\cite{AKLT1,AKLT2} Sutherland, \cite{Sutherland} Kl\"umper,
\cite{Klumper1,Klumper2,Klumper3} Barber \cite{Barber} which supported Haldane's hypothesis
and established an interesting phase-diagram\cite{Schollwoeck,FathSolyom1} as shown in Fig.~\ref{phasediagram}.
Experimental results on quasi-one-dimensional spin-1 compounds such as "NENP",
$\rm CaNiCl_3$ or $\rm AgVP_2S_6$ also confirmed the gap and the existence of
effective spin-$\frac{1}{2}$ spins near boundaries.\cite{Armstrong,Mutka}
The spin-1 chain was also one of the driving forces in the development of
the density matrix renormalization group (DMRG) algorithm which in turn
provided  very accurate estimates of the excitation spectrum and the correlation lengths
at the Heisenberg point.\cite{White1,White2}

By changing the ratio of the Heisenberg coupling $J_1$ and biquadratic exchange term $J_2$,
the system can be tuned through at least three antiferromagnetic regions
and one ferromagnetic phase as shown in Fig.~\ref{phasediagram}.
The
antiferromagnetic phase consists of three regions, Trimer, Haldane, and Dimer, of which
the last two are gapped.
Because in most substances the biquadratic exchange term is much smaller
compared to the bilinear term, the experimental realization in regions with
dominant biquadratic exchange term was not possible for a long time. First
experimental success was achieved with $\rm LiVGe_2O_2$ which appears to be well
described by a large positive value of $J_2$.\cite{Millet}

The similarity of the phase diagrams of the spin-1 chain compared to the spin-$\frac{1}{2}$
chain with next nearest
neighbor coupling is striking. In particular, the spin-$\frac{1}{2}$ chain also shows 
three antiferromagnetic regions, two of which are believed to be gapped and one
ferromagnetic phase as shown in Fig.~\ref{phasediagram}.  It is known that
the AKLT point in the spin-1 chain is in the same phase as the Majumdar-Ghosh
point\cite{MajumdarGhosh} in the spin-$\frac{1}{2}$ chain,
i.e.~the two points $B$ and $b$ in Fig.~\ref{phasediagram} can be connected
in a more general parameter space.\cite{Mikeska}   We also see that the
two gapped phases in both the spin-$1$ and the spin-$\frac{1}{2}$ chain are separated by
an integrable point with SU(2)$_2$ symmetry (points $G,\ g$).  Moreover, in both cases
the phase transition
to the ferromagnetic behavior occurs at points where the Hamiltonian
can be written in terms of
total spin projection operators of neighboring spins (points $D,d,E,e$).
Therefore, the main difference between the two phase diagrams is that 
in the spin-1 chain the isotropic point $A$ happens to be in a gapped phase,
while the isotropic point $a$ falls in a gapless region in the spin-$\frac{1}{2}$ chain.

In this paper, we develop a highest weight state basis which can be used to calculate 
a number of exact properties, especially at the ''projection points'' $B$, $D$ and $E$, 
where the Hamiltonian can be written as a sum over singlet, triplet, or quintet 
projection operators of two neighboring
spins, respectively.
In section \ref{basis}  we introduce the general total spin basis, which is in spirit
similar to the valence bond basis for spin-$\frac{1}{2}$ systems where the total spin of
pairs of spins in the system is specified.   
In section \ref{ferro}, we demonstrate how
to calculate the ferromagnetic excitations with total spin $s=N-1$
in the new basis as a simple illustration.
At the singlet projection point $D$, the ferromagnetic excitations become dispersionless
and it is possible to construct $s=N-2$ excitations in agreement with
earlier results\cite{Parkinson} as shown in section \ref{P0}.
In section \ref{P1}, we consider the triplet projection 
point $E$ at $J_1=-J_2<0$ ($\Theta=\frac{3\pi}{4}$) in a system with an even number of 
sites $N$.  
At this phase transition the 
antiferromagnetic $s=0$  and ferromagnetic $s=N$ ground states are degenerate, 
but also 
ground states with any even total spin $s$ exist, 
leading to a large degeneracy.
We are able to calculate the correlation functions exactly which remarkably do not decay 
along the chain, even for the antiferromagnetic ground state.
Using the highest weight state basis, excited states can also be constructed exactly 
at this point.
For completeness we also show how to express the AKLT state at the
quintet projection point $B$ ($\Theta=-\arctan 1/3$) in the new
basis in section \ref{P2}.
We conclude in section \ref{conclusions}.  The appendix explains how operators 
are applied
and scalar products are computed in the highest weight state basis.

\section{The total spin basis for spin-1 systems} 
\label{basis}
In this section a basis set of highest weight states
with given total spin $s$ is introduced.  Similar to the 
valence bond basis for spin-$\frac{1}{2}$ systems, it is possible to specify pairs of 
spins (so-called valence bonds) which have definite total spin. 
In spin-1 systems, pairs of spins 
can now be quintets ($s=2$), triplets ($s=1$) or
singlets ($s=0$).  The most important difference to an ordinary $S^z$-basis is
that in this construction ''bonds'' are specified instead of local quantum numbers.

We use  a notation similar to the conventional
one for the spin-$\frac{1}{2}$ chain by Majumdar and Ghosh\cite{MajumdarGhosh} 
$\{i_1,i_2\}_{m=0}\equiv \ket{s=0,m=0}$, $[i_1,i_2]_m\equiv \ket{s=1,m}$ 
and $(i_1,i_2)_m\equiv \ket{s=2,m}$, where $s$ is the total spin 
of the pair of spins at sites $i_1$, $i_2$ and $m$ the $S^z$ eigenvalue. 
An important simplification for $SU(2)$ invariant systems is that 
only highest weight states (hws) with $m=s$ 
have to be considered as representatives of 
degenerate multiplets, since
it is possible to construct an entire degenerate multiplet by applying 
the total spin lowering operator ${\cal S}^-=\sum_i S_i^-$, which commutes with $H$. 

Accordingly, it is useful to define 
''bonds'' as pairs of two spins, which are in a 
highest weight state of total spin $2$, $1$, or $0$
\begin{eqnarray}\label{notation}
(i_1,i_2)_2  &\equiv& (i_1,i_2)= \ket{++} \nonumber \\
{[i_1,i_2]_1}  &\equiv& [i_1,i_2]= (\ket{+0}-\ket{0+})/ \sqrt{2}  \nonumber \\
\{i_1,i_2\}_0 &\equiv& \{i_1,i_2\}=(\ket{+-}+\ket{-+}-\ket{00})/ \sqrt{3}\ .
\end{eqnarray} 
Singlet and quintet bonds have even parity under exchange of indices, 
while the triplet bonds are odd.
Using this notation it is now possible to construct a highest weight state
of the total system by specifying all bonds as follows
\begin{equation}\label{hws} \ket{\psi_{s}(i_1,...,i_N)}=
\overbrace{(i_1,i_2)...(.,.)}^{Q}\overbrace{[.,.]}^{T}
\overbrace{\{.,.\}...\{i_{N-1},i_N\}}^{S}\ ,
\end{equation}
assuming an even number of lattice sites $N$. 
The number of quintet-, triplet- and singlet-bonds, $Q$, $T$ and $S$ defines
the total spin $s=2Q+T$.  The wave-function is a function of the  
pairwise different ordered indices $i_1,i_2,...,i_N$, which are used to specify  
the bonds.  The order of the quintet indices 
does not matter since 
all spins in quintet-bonds are in the state $|+\rangle$, so Eq.~(\ref{hws}) can be 
simplified by only specifying 
which spins belong to the quintets $(i_1,... , i_{2Q})$.    
Also the order within each singlet and triplet bracket 
is irrelevant up to a possible minus sign.
To simplify the calculations it is sometimes useful  
to specify the bonds graphically as follows
\begin{equation}\label{graphical_hws}
\{i_1,i_2\}=\put(-2,-10){$i_1$}\qbezier(0,0)(10,18)(20,0)\put(18,-10){$i_2$} \qquad \qquad [i_1,i_2]=\put(-2,-10){$i_1$}\line(0,1){10}\put(0,10){\vector(1,0){20}}\put(16,0){ \line(0,1){10}}\put(18,-10){$i_2$}  \qquad \qquad (i_1,i_2)=\ \put(3,-10){$i_1$}\bullet \quad \put(3,-10){$i_2$}\bullet\ ,
\end{equation}
where no bond connections are assigned to 
quintets. 

Clearly, states of the form (\ref{hws}) are indeed highest weight states of
the system, since
\begin{eqnarray}\label{properties}
S^z_{\rm{tot}}\ket{\psi_{s}}&=&m\ket{\psi_{s}} \nonumber \\
S^2_{\rm{tot}}\ket{\psi_{s}}&=&s(s+1)\ket{\psi_{s}}\ ,
\end{eqnarray}
where $s=m=2 Q + T$.  Here, the first line is obvious from Eq.~(\ref{notation}) 
and the second line follows since the total spin $s$ has to be at least 
$m$, but can be at most the sum of the individual bond spins $s=2Q + T$.

States of the form (\ref{hws}) can now be used as basis states in order 
to express any hws of the system.  For this purpose it is sufficient to
only consider basis states with no triplets $T=0$ (for $s=2Q$ even) or
one triplet pair $T=1$ (for $s=2Q+1$ odd), since states with $T>1$ can be expressed by 
the following linear combination
\begin{widetext}
\begin{eqnarray}\label{decomposition}
 [1,2][3,4]  \propto &(2,3)\{1,4\} &+(1,4)\{2,3\}  -  
(1,3)\{2,4\}  -  (2,4)\{1,3\} \nonumber \\
&  &
\put(-125,0){\line(0,1){10}\put(0,10){\vector(1,0){20}}\put(16,0){ \line(0,1){10}}
\put(25,0){\line(0,1){10}\put(0,10){\vector(1,0){20}}\put(16,0){ \line(0,1){10}}}} 
\put(-75,0){$\propto$\put(8,0){\qbezier(0,0)(25,22)(50,0)\put(13,0){$\bullet$\put(10,0){$\bullet$}}}}
\put(0,0){+ \put(5,0){$\bullet$\put(30,0){$\bullet$}\put(5,0){\qbezier(0,0)(10,22)(20,0)}}}
\put(65,0){--- \put(10,0){$\bullet$\put(20,0){$\bullet$}\put(8,0){\qbezier(0,0)(14,22)(28,0)}}}
\put(137,0){--- \put(10,0){\qbezier(0,0)(14,22)(28,0)\put(12,0){$\bullet$
\put(20,0){$\bullet$}}}\put(55,0){.}}
\end{eqnarray}

In the hws basis defined in this way,  
it is possible to determine the action of any $SU(2)$ invariant 
operator involving two spins, i.e.~linear combinations of 
$\textbf{S}_i\cdot \textbf{S}_{i+1}$ and 
$(\textbf{S}_i\cdot \textbf{S}_{i+1})^2$.  For example for states 
with two singlet bonds we find the following relations
\begin{eqnarray}\label{operators}
\textbf{S}_i\cdot \textbf{S}_{i+1}\{j,i\}\{i+1,k\}&=&\{j,i+1\}\{i,k\}
-\{j,k\}\{i,i+1\}  \nonumber \\
\left( \textbf{S}_i\cdot \textbf{S}_{i+1}\right)^2\{j,i\}\{i+1,k\}&=&\{j,i\}\{i+1,k\}
+\{j,k\}\{i,i+1\}\ ,  
\end{eqnarray}
or equivalently
\begin{eqnarray}
\put(-125,0){$\textbf{S}_i\cdot \textbf{S}_{i+1}$}
\put(-80,0){\put(-2,-10){$j$}\qbezier(0,0)(10,22)(20,0)\put(18,-10){$i$}
\put(40,0){\put(-10,-10){$i+1$}\qbezier(0,0)(10,22)(20,0)\put(18,-10){$k$}}}
\put(-4,0){=}
\put(20,0){\put(-2,-10){$j$}\qbezier(0,0)(20,22)(40,0)\put(28,-10){$i+1$}
\put(20,0){\put(-2,-10){$i$}\qbezier(0,0)(20,22)(40,0)\put(38,-10){$k$}}}
\put(90,0){---}
\put(110,0){\put(-2,-10){$j$}\qbezier(0,0)(29,22)(58,0)\put(56,-10){$k$}
\put(20,0){\put(-2,-10){$i$}\qbezier(0,0)(10,16)(20,0)\put(8,-10){$i+1$}}}
\nonumber \\
\put(-140,0){$(\textbf{S}_i\cdot \textbf{S}_{i+1})^2$}
\put(-80,0){\put(-2,-10){$j$}\qbezier(0,0)(10,22)(20,0)\put(18,-10){$i$}
\put(40,0){\put(-10,-10){$i+1$}\qbezier(0,0)(10,22)(20,0)\put(18,-10){$k$}}}
\put(-4,0){=}
\put(20,0){\put(-2,-10){$j$}\qbezier(0,0)(10,22)(20,0)\put(18,-10){$i$}
\put(40,0){\put(-10,-10){$i+1$}\qbezier(0,0)(10,22)(20,0)\put(18,-10){$k$}}}
\put(91,0){+}
\put(110,0){\put(-2,-10){$j$}\qbezier(0,0)(29,22)(58,0)\put(56,-10){$k$}
\put(20,0){\put(-2,-10){$i$}\qbezier(0,0)(10,16)(20,0)\put(8,-10){$i+1$}}}
\put(190,0){.}
\end{eqnarray}
\end{widetext}
Since the terms in the Hamiltonian (\ref{model}) operate only on two spins at a time, it is
always sufficient to consider clusters  of two bonds irrespective of the length of
the chain. A complete list how $\textbf{S}_i\cdot \textbf{S}_{i+1}$ and
$(\textbf{S}_i\cdot \textbf{S}_{i+1})^2$ operate on the states can be found 
in appendix \ref{relation}. 

{}From those relations (\ref{relation})
it is clear that the representation of linear combinations
in the hws basis 
in Eq.~(\ref{hws}) is closed under the operation of any local $SU(2)$ operator,
including permutations.  Therefore, starting with any hws all other states in the
corresponding sector of the Hilbert space can be represented 
as a linear combination 
in the hws basis (\ref{hws}).  However, the linear combinations are not necessarily unique
since the hws basis is overcomplete and not orthogonal.
In particular, for the $s=0$ sector the hws basis is linearly independent up to 
$N=6$ spin, while for $N=8$ the basis becomes overcomplete.
This means that no relation between singlet states with 3 singlet bonds exists, 
while there are 14 (relatively complicated) relations involving 4 singlet bonds, which 
will not be discussed here.  
In general, there are $N!/(2^{(N-s)/2}(\frac{N-s}{2})!s!)$ basis states of the form
(\ref{hws}) for a given even spin $s=2Q$.
For comparison, if only the quantum number in the $z$-direction $S^z=0$
in the ordinary $S^z$ basis is specified, the number of basis states 
is given by $1+\sum_{N_+=1}^{N/2} \frac{N!}{(N-2N_+)!N_+!N_+!}$
which is a larger basis than the $s=0$ hws basis up to $N=18$.
The situation for $N=4$ spins is discussed in detail in appendix \ref{4spins} as an example.

It is also possible to determine scalar products in the hws basis using a straight-forward
algorithm 
as presented in appendix \ref{product}.
In particular, the scalar product between two total spin $s=0$
states can be obtained by the minimum number $\gamma$ of index exchanges $V_{i_k,i_l}$
needed to transform the indices $i_1,...,i_N$ into an equivalent bond configuration 
corresponding to indices $i_1',...,i_N'$
\begin{eqnarray}\label{scalarproduct}
& & \langle \psi_{s=0}(i_1,...,i_N)|\psi_{s=0}(i_1',...,i_N')\rangle = \\
& & \ \ \ \ \ \ \ 
\langle\psi_{s=0}(i_1,...,i_N) | \overbrace{V...V}^{\gamma}|\psi_{s=0}(i_1,...,i_N)\rangle 
=\frac{1}{3^{\gamma}}\ . \nonumber
\end{eqnarray}
Interestingly, this means that any basis state in the $s=0$ sector is
never orthogonal to any other $s=0$ basis state.

\section{Ferromagnetic excitations} \label{ferro}
In order to demonstrate how to apply the new basis, we consider as an example
excitations on the 
ferromagnetic state $\ket{F}=(1,2,...,N)$ which 
is always an eigenstate with 
 energy $E_0=N(J_1-J_2)=JN(\cos \Theta-\sin \Theta)$. 
In order to construct a spin-wave excitation $\ket{k}$ with total spin $s=N-1$
we can write
\begin{equation}
\ket{k}= \sum_{i_1\neq i_2}e^{\imath k i_1}[i_1,i_2](i_3,...,i_N)\ . \label{k}
\end{equation}
Here, we require $k\neq 0$ since the $k=0$ ''excitation'' 
of the ordinary spin-wave construction 
actually corresponds to a $s=N$ multiplet state.
We will now show explicitly that the state in Eq.~(\ref{k}) is 
an eigenstate of the Hamiltonian (\ref{model}) for any $J_1$ and $J_2$.
In order to apply the 
operators $\textbf{S}_i\cdot \textbf{S}_{i+1}$ and $(\textbf{S}_i\cdot \textbf{S}_{i+1})^2$, 
we write states involving triplet bonds with the sites $i$  or $i+1$ separately
\begin{eqnarray}
\ket{k}&=& e^{\imath k i}[i,i+1](...)+e^{\imath k(i+1)}[i+1,i](...) \nonumber \\
&&+\sum_{i_3\neq i+1}e^{\imath k i}[i,i_3](..,i+1,..)\nonumber \\
&& +\sum_{i_3\neq i}e^{\imath k (i+1)}[i+1,i_3](..,i,..) \nonumber \\
&&+\sum_{i_3\neq i+1}e^{\imath k i_3}[i_3,i](..,i+1,..)\nonumber \\
& & +\sum_{i_3\neq i}e^{\imath k i_3}[i_3,i+1](..,i,..) \nonumber \\
&&+\sum_{i_4,i_3\neq i,i+1}e^{\imath k i_3}[i_3,i_4](..,i,i+1,..)\ . \label{decompose}
\end{eqnarray}
Now we use the relation from appendix \ref{relation}
\begin{eqnarray}
\textbf{S}_i\cdot \textbf{S}_{i+1}(j,i)[i+1,k]& =& (  j,i+1)[i,k] \nonumber \\
(\textbf{S}_i\cdot \textbf{S}_{i+1})^2(j,i)[i+1,k]& =& (j,i)[i+1,k]\ .  \label{sop1}
\end{eqnarray}
By summing over all $i$, we obtain 
\begin{eqnarray}
\sum_{i=1}^N\textbf{S}_i\cdot \textbf{S}_{i+1}\ket{k}&=&N\ket{k}+2(\cos(k)-1)\ket{k} \nonumber \\
\sum_{i=1}^N(\textbf{S}_i\cdot \textbf{S}_{i+1})^2&=&N\ket{k}\ .
\end{eqnarray}
This means that for a general Hamiltonian (\ref{model})
\begin{equation}
H\ket{F}=E_0\ket{F} \qquad H\ket{k}=E_0\ket{k}+2J_1(\cos(k)-1)\ket{k}\ , \label{f-disp}
\end{equation}
which is the well-known ferromagnetic dispersion.

\section{Model at $J_1=0>J_2$ ($\Theta=-\frac{\pi}{2}$)}
\label{P0}
One immediate consequence of the dispersion relation (\ref{f-disp})
is that the excitations become dispersionless at the phase transition point 
$D$ for $J_1=0$
\begin{equation}\label{modelB}
H=-J_2\sum_{i=1}^N (\textbf{S}_i\cdot \textbf{S}_{i+1})^2\ , 
\end{equation}
where we assume $J_2<0$. 
At this point, it is also possible to express the Hamiltonian in terms of a
total spin $s=0$ projection operator of two neighboring spins
$P_0(\textbf{S}_i,\textbf{S}_{i+1})=-\frac{1}{3}+\frac{1}{3}(\textbf{S}_i\cdot
\textbf{S}_{i+1})^2$
\begin{equation}
H=E_0-3J_2\sum_{i=1}^N P_0(\textbf{S}_i,\textbf{S}_{i+1})\ ,
\end{equation}
where $E_0=-NJ_2$. For this model we are able to compare with known results,\cite{Parkinson} which 
serves as an illustration of how to use the hws basis. 
We know from Eq.~(\ref{sop1}) that {\it any} state with total spin $s=N-1$ 
is an eigenstate and degenerate with the ground state.  
Excitations of the form $\ket{k}=\sum_{i_1} \exp(\imath k i_1)\{i_1,i_1+1\}(i_3,i_4...i_N)$ with total spin $s=N-2$ also become eigenstates at this point. 
This can be shown again by decomposing $\ket{k}$ in the following form 
\begin{eqnarray}\label{expantionA}
\ket{k} &=& \sum_{i_1=1}^N e^{\imath k i_1}\{i_1,i_1+1\}(.,..,.)  \\
	&=&e^{\imath k i}\{i,i+1\}(.,..,.)+e^{\imath k(i-1)}\{i-1,i\}(.,..,.)\nonumber \\
	&&+e^{\imath k(i+1)}\{i+1,i+2\}(.,..,.) \nonumber \\
	&&+\sum_{i_3\neq i-1,i,i+1}e^{\imath k i_3}\{i_3,i_3+1\}(..,i,i+1,..) \ . \nonumber
\end{eqnarray}
Using Eq. (\ref{relations}) from the appendix (\ref{relation}), we find
\begin{eqnarray}
(\textbf{S}_i \cdot \textbf{S}_{i+1})^2 \ket{k}&    =& 
3 e^{\imath k i}\{i,i+1\}(.,..,.)  \\
& & +[e^{\imath k}+e^{-\imath k}]e^{\imath k i}\{i,i+1\}(.,..,.)+\ket{k}\ . \nonumber
\end{eqnarray}
Finally, we sum over all $i$, and obtain the dispersion relation
$E_2(k)=-J_2[3+2\cos(k)]+E_0$, which is gapped in agreement with earlier results from
Ref.~[\onlinecite{Parkinson}].

\section{Model at $J_1=-J_2<0$ ($\Theta=\frac{3\pi}{4}$)}
\label{P1}
We now turn to the other ferro-antiferromagnetic phase 
transition point $E$ at $J_1=-J_2<0$. Using the 
basis of hws in Eq.~(\ref{hws}) it is not only possible to 
describe all degenerate ferro- and antiferromagnetic ground states, but also 
to calculate correlation functions and excited states exactly.
The model is now given by the Hamiltonian 
\begin{equation}\label{modelA}
H=J_1 \sum_{i=1}^N \left( \textbf{S}_i\cdot \textbf{S}_{i+1}+(\textbf{S}_i\cdot
\textbf{S}_{i+1})^2\right)\ , 
\end{equation}
for a periodic chain with even number of sites $N$.
It is possible to express the Hamiltonian (\ref{modelA})
in terms of triplet projection operator of two neighboring spins 
$P_1(\textbf{S}_i,\textbf{S}_{i+1})=(1-\textbf{S}_i\cdot
\textbf{S}_{i+1}/2-(\textbf{S}_i\cdot \textbf{S}_{i+1})^2/2)$
\begin{equation}
H=E_0-2J_1\sum_{i=1}^N P_1(\textbf{S}_i,\textbf{S}_{i+1})\ ,
\end{equation}
where $E_0=2NJ_1$. Since the operator $P_1$ can only have positive eigenvalues and $J_1<0$, 
the system has so-called optimum ground states,\cite{klumper4} which must 
be in a configuration where no two 
neighboring spins are in a triplet configuration, so that $P_1 |0\rangle=0$ for
all nearest neighbors. 

Another useful representation can be obtained by using the 
operator $V_{i,i+1}$, which exchanges the quantum numbers of two neighboring sites.
Since both singlets and quintets are even under exchange, we have
$P_1(\textbf{S}_i,\textbf{S}_{i+1})=(1-V_{i,i+1})/2$ and therefore the
Hamiltonian becomes
\begin{equation}\label{hamilton}
H=J_1N+J_1\sum_{i=1}^N V_{i,i+1}\ ,
\end{equation}
which is in fact $SU(3)$ invariant.\cite{Sutherland,FathSolyom1}
Hence, ground states of the Hamiltonian (\ref{hamilton}) can be 
constructed by requiring  
$V_{i,i+1}\ket{0}=\ket{0}$ for all $i$, i.e.~the ground
states are invariant under application of all permutation operators
$V_{i,i+1}$.\cite{FathSolyom1,Chubukov} In order to
construct such a ground state, we use the hws notation (\ref{hws}) to 
describe the antiferromagnetic and the ferromagnetic ground state
as follows
\begin{eqnarray}\label{groundstate}
\ket{F}  &= & (1,2,...,N-1,N) 
\nonumber \\
\ket{AF} &\propto & \sum_{\rm{P}}\{i_1,i_2\}\{i_3,i_4\}\{i_5,i_6\}...\ ,
\end{eqnarray}
where $\sum_P$ is the sum over all possible permutations of $i_1, ..., i_N$. 
It is clear, that the
states defined in this way are eigenstates of all permutations $V_{i,i+1}$
and therefore
immediately ground states of the Hamiltonian (\ref{modelA}) with definite total 
spin $s=N$ and $s=0$, respectively.   Therefore, the energies of the 
ferro- and antiferromagnetic states cross at the phase transition point 
as expected, but in addition 
we find that it is possible to construct ground states with
any even spin $s$ by simply combining permutations of quintet and
singlet bonds
\begin{equation}
\ket{0}_s \propto 
\sum_P(i_1,i_2,...,i_{s-1},i_{s})\{i_{s+1},i_{s+2}\}...\{i_{N-1},i_N\} \label{0s}\ ,
\end{equation}
which is in accordance with $SU(3)$ invariance.  There are no other ground states.

It is interesting to note that in the spin-$\frac{1}{2}$ chain at the phase transition 
point $e$ in 
Fig.~\ref{phasediagram} the ground state can also be written 
as a permutation over all possible valance bonds.\cite{Natsume}

For each ground state all pairs of spins are equally entangled with each other. 
Consequentially, also the correlation function is the same between any
two spins in the chain, independent of distance
\begin{equation}
{\phantom\rangle}_s\!\bra{0} \textbf{S}_i \cdot \textbf{S}_j \ket{0}_s = \frac{s (s+1)-2 N}{N(N-1)}\ , 
\end{equation}
which follows from expanding 
$\langle \textbf{S}_{\rm{tot}}^2 \rangle=\langle \sum_{i \neq j}\textbf{S}_i
\cdot \textbf{S}_j+\sum_i \textbf{S}_i^2\rangle$ and can also be verified by direct 
calculation with help of the scalar product in the appendix \ref{product}.
The correlation function varies from 
$\bra{AF} \textbf{S}_i \cdot \textbf{S}_j \ket{AF} = -\frac{2}{N-1}$ up to 
$\bra{F} \textbf{S}_i \cdot \textbf{S}_j \ket{F}=1$. 
Higher order correlation functions can then also be determined iteratively by the use of
$P_1(\textbf{S}_i,\textbf{S}_j)\ket{0}_s=\left(1-\textbf{S}_i\cdot
\textbf{S}_j/2-(\textbf{S}_i\cdot \textbf{S}_j)^2/2\right) \ket{0}_s = 0$ for any pair 
of spins, so that e.g. $\langle (\textbf{S}_i \cdot
\textbf{S}_j)^2\rangle =2 -\langle \textbf{S}_i \cdot
\textbf{S}_j\rangle$.

A magnon excitation can now be constructed in a similar spirit as in Eq.~(\ref{k}) on any
of the ground states.  For example a spin-$1$ excitation on the antiferromagnetic state
is given by
\begin{eqnarray}\label{states}
\ket{k}_{\rm{AF}} &= & \sum_{i_1\neq
i_2}e^{\imath k i_1}[i_1,i_2]\sum_{\rm{P'}}\{i_3,i_4\}...\ ,
\end{eqnarray} 
with $k\neq 0$. 
In order to show that this state is an eigenstates of the Hamiltonian (\ref{hamilton}),
a decomposition as in Eq.~(\ref{decompose}) can be used.
Applying $V_{i,i+1}$ and then summing again over all
$i$, we obtain 
\begin{equation}
H\ket{k}_s=E_1(k)\ket{k}_s \qquad E_1(k)=2J_1(\cos(k)-1)+E_0 \ ,
\end{equation}
where $\ket{k}_s$ is now the corresponding excitation 
on the ground state $\ket{0}_s$ with any even spin $s$
in Eq.~(\ref{0s}).

Interestingly, the permutational 
ground states in Eq.~(\ref{0s}) 
are in fact independent of the individual coupling strengths along the chain 
and are even ground states of higher dimensional 
Hamiltonians as long as $J_1=-J_2$.  
However, 
the excited states in Eq.~(\ref{states}) require  translational 
invariance.

\section{Model at $J_1=-J_2/3>0$ ($\Theta=-\arctan{\frac{1}{3}}$)}
\label{P2}
Finally we consider the famous AKLT model\cite{AKLT1,AKLT2}
at point $B$ in the phase diagram
\begin{eqnarray}\label{modelC}
H&=&J_1\sum_{i=1}^N\left(\textbf{S}_i\cdot
\textbf{S}_{i+1}+\frac{1}{3}(\textbf{S}_i\cdot
\textbf{S}_{i+1})^2\right)\nonumber \\
&=&E_0+2J_1\sum_{i=1}^N P_2(\textbf{S}_i,\textbf{S}_{i+1}),
\end{eqnarray}
where $J_1>0$, $E_0=-\frac{2}{3}NJ_1$ and 
$P_2(\textbf{S}_i,\textbf{S}_{i+1})=1/3+\textbf{S}_i\cdot
\textbf{S}_{i+1}/2+(\textbf{S}_i\cdot \textbf{S}_{i+1})^2/6$ is the projection operator 
onto the quintet state of two neighboring spins.
It is of course well known how to express the ground state $\ket{0}$
at this point,\cite{AKLT1,AKLT2}
but it is instructive to gain an alternative description in terms  
of the hws basis. 

Analogously to the other projection points, the Hamiltonian (\ref{modelC})
has an optimum ground state $\ket{0}$, which must obey
$P_2(\textbf{S}_i,\textbf{S}_{i+1})\ket{0}= 0$ for all adjacent spins. 
Using the relations in appendix \ref{relation} we know 
\begin{eqnarray}\label{projectionrelation}
&P_2(\textbf{S}_i,\textbf{S}_{i+1})&\{i,i+1\}=0 \nonumber \\
&P_2(\textbf{S}_i,\textbf{S}_{i+1})&\left(\{j,i\}\{i+1,k\}-\{j,i+1\}\{i,
k\}\right)=0\  \nonumber \\
&P_2(\textbf{S}_i,\textbf{S}_{i+1})& (
\put(5,0){\qbezier(0,0)(10,22)(20,0)\put(30,0){\qbezier(0,0)(10,22)(20,0)}}
\put(62,0){---}
\put(80,0){\qbezier(0,0)(14,22)(28,0)\put(12,0){\qbezier(0,0)(14,22)(28,0)}}
\put(125,0){)=0.}
\end{eqnarray}
Accordingly, we can construct the ground state by ensuring that for neighboring 
spins which
are not coupled by a singlet bond the corresponding crossed state is subtracted.
The following state is therefore a ground state
\begin{equation}\label{groundstatevbs}
\ket{0}\propto \sum_P (-1)^{\# crossings(P)}\{1,2\}\{3,4\}...\{N-1,N\}\ ,
\end{equation}
where the number of crossings  is defined as the crossing of singlet bonds
in the graphical representation in Eq.~(\ref{graphical_hws}).  For example 
in the $N=4$ chain the ground state is given by  
\begin{eqnarray}
\ket{0}	&\propto& \left(\{1,2\}\{3,4\}+\{1,4\}\{2,3\}-\{2,4\}\{1,3\}\right) \nonumber \\
	&\propto& \put(5,0){\qbezier(0,0)(10,22)(20,0)\put(27,0){\qbezier(0,0)(10,22)(20,0)}}
\put(61,0){+}
\put(77,0){\qbezier(0,0)(24,22)(48,0)\put(14,0){\qbezier(0,0)(10,16)(20,0)}}
\put(130,0){---}
\put(146,0){\qbezier(0,0)(14,22)(28,0)\put(12,0){\qbezier(0,0)(14,22)(28,0)}}
\put(190,0){.}  \nonumber
\end{eqnarray}

For a finite chain with open boundary conditions, the Hamiltonian (\ref{modelC})
has a four degenerate ground states, one singlet and one triplet.  The singlet
given by Eq.~(\ref{groundstatevbs}) with $s=0$. 
By using the relation 
from the appendix \ref{relation} we find for triplet states
\begin{eqnarray}\label{projectionrelation2}
&P_2(\textbf{S}_i,\textbf{S}_{i+1})&[i,i+1]=0  \\
&P_2(\textbf{S}_i,\textbf{S}_{i+1})&\left([j,i]\{i+1,k\}-[j,i+1]\{i,k\}
\right)=0\ .
\nonumber
\end{eqnarray} 
Hence, the triplet ground state is analogously given by  
\begin{equation}
\ket{0}= \sum_{P'} (-1)^{\# crossings(P)}[1,2]\{3,4\}...\{N-1,N\}\ ,
\end{equation}
where the permutations are restricted so that 
the indices in the triplet bond always remain in ascending order.
This ground state is a triplet of
 total spin $s=1$ in accordance with previous results.\cite{AKLT1,AKLT2}
\section{Conclusions}
\label{conclusions}
We have introduced a highest weight state basis as a total spin
representation  for the Hilbert space of
 spin-1 systems.  In this basis it is possible to compute scalar products and
the action of $SU(2)$ invariant operators for analytical and
numerical calculations.  However, the new basis states
are not orthogonal and overcomplete.

In the ferromagnetic phase and at 
the phase transition point $D$ ($\Theta=-\pi/2$) it is possible to construct 
spin wave excitations explicitly in the new basis.  Also the AKLT ground states
at point $B$ ($\tan \Theta=- 1/3$) can be explicitly expressed using the 
highest weight states.

At the phase transition point $E$ ($\Theta=3\pi/4$) we find all degenerate ground states 
which are multiplets of even total spin $s=0,2,...,N$.  The states
can be  expressed as permutations over all hws bonds. 
It is possible to determine the spin correlation functions 
$\langle \textbf{S}_i\cdot \textbf{S}_j \rangle$ which do not decay along the
chain, even for the antiferromagnetic state.
Corresponding excited states were also determined in the hws basis.
There is some hope that these results  at point $E$ 
can be used in future works as an ansatz for
wave-functions in the antiferromagnetic 
region in order to investigate the phase between points $E$ and $F$, which is still
not fully understood.

\begin{acknowledgments}
We are grateful for useful discussions with Alexander Struck. 
This work was supported in part by the Graduate Class of Excellence MATCOR
funded by the State of Rheinland-Pfalz, Germany.
\end{acknowledgments}

\appendix\

\section{Action of $SU(2)$ invariant operators} \label{relation}
The action of $h_{i,i+1}=\textbf{S}_i\cdot \textbf{S}_{i+1}$ on bonds where both indices are in the same 
singlet, triplet,  or quintet coupling is immediately given by the respective eigenvalues
\begin{eqnarray}
h_{i,i+1}
\{i,i+1\}&=&-2\{i,i+1\} \nonumber \\
h_{i,i+1}
[i,i+1]&=&-[i,i+1] \nonumber \\
h_{i,i+1}
(i,i+1)&=&(i,i+1)\ . 
\end{eqnarray}
Using the definition in Eq.~(\ref{notation}) 
it is straight-forward to determine the action of 
$h_{i,i+1}=\textbf{S}_i\cdot \textbf{S}_{i+1}$ on all possible hws states as
follows
\begin{eqnarray}
h_{i,i+1}
\{j,i\}\{i+1,k\}&=&\{j,i+1\}\{i,k\}-\{j,k\}\{i,i+1\} \nonumber \\
h_{i,i+1}
[j,i]\{i+1,k\}&=&[j,i+1]\{i,k\}-[j,k]\{i,i+1\}
\nonumber \\
h_{i,i+1}
(j,i)\{i+1,k\}&=&(j,i+1)\{i,k\}-(j,k)\{i,i+1\}
\nonumber \\
h_{i,i+1}
(j,i)[i+1,k]&=&(j,i+1)[i,k]\ .  
\label{relations}
\end{eqnarray}
The application of $h_{i,i+1}^2=(\textbf{S}_i\cdot \textbf{S}_{i+1})^2$ is then given by applying 
$\textbf{S}_i\cdot \textbf{S}_{i+1}$ twice 
\begin{eqnarray}
h_{i,i+1}^2
\{j,i\}\{i+1,k\}&=&\{j,i\}\{i+1,k\}+\{j,k\}\{i,i+1\} \nonumber \\
h_{i,i+1}^2
[j,i]\{i+1,k\}&=&[j,i]\{i+1,k\}+[j,k]\{i,i+1\}
\nonumber \\
h_{i,i+1}^2
(j,i)\{i+1,k\}&=&(j,i)\{i+1,k\}+(j,k)\{i,i+1\}
\nonumber \\
h_{i,i+1}^2
(j,i)[i+1,k]&=&(j,i)[i+1,k]\ .  
\end{eqnarray}
In this way the action of 
any $SU(2)$ invariant operator involving two spins
can be calculated as a linear combination of $\textbf{S}_i\cdot \textbf{S}_{i+1}$ and
$(\textbf{S}_i\cdot \textbf{S}_{i+1})^2$ using above relations.

\section{HWS basis for $N=4$ spins} \label{4spins}
We consider a spin-$1$ system consisting of $4$ sites 
and find all hws states of the form (\ref{hws}), using at most one triplet bond.
In the spin $s=0$ sector we find
\begin{eqnarray}\label{4spin1chain}
 & & \{1,2\}\{3,4\},\{1,3\}\{2,4\},\{1,4\}\{2,3\}; 
\end{eqnarray}
in the spin $s=1$ sector we find
\begin{eqnarray}
 & & {[1,2]}\{3,4\},[1,3]\{2,4\},[1,4]\{2,3\},\nonumber \\ 
 & & {[3,4]}\{1,2\},[2,4]\{1,3\},[2,3]\{1,4\}; 
\end{eqnarray}
in the spin $s=2$ sector we find
\begin{eqnarray}
 & & (1,2)\{3,4\},(1,3)\{2,4\},(1,4)\{2,3\}, \nonumber \\ 
 & & (3,4)\{1,2\},(2,4)\{1,3\},(2,3)\{1,4\}; 
\end{eqnarray}
in the spin $s=3$ sector we find
\begin{eqnarray}
 & & (1,2)[3,4],(1,3)[2,4],(1,4)[2,3],\nonumber \\ 
 & & (3,4)[1,2],(2,4)[1,3],(2,3)[1,4]; 
\end{eqnarray}
and the ferromagnetic state is
\begin{eqnarray}
 & & (1,2,3,4) \ .
\end{eqnarray}

Using the addition rules for angular momenta 
$\textbf{1}\otimes\textbf{1}\otimes\textbf{1}\otimes\textbf{1}=3 \cdot
\textbf{0}\ \oplus\  6\cdot \textbf{1}\ \oplus\  6\cdot \textbf{2}\ \oplus\  3\cdot
\textbf{3}\ \oplus\  1\cdot \textbf{4}$ we see that the states of spin $s=0$, 
$s=1$, $s=2$ and $s=4$ are linearly independent, as can also be checked directly using
Eq.~(\ref{notation}).  For the $s=3$ sector we obtain the following linear relations
\begin{eqnarray}\label{example}
\put(-100,0){
\line(0,1){10}\put(0,10){\vector(1,0){20}}\put(16,0){ \line(0,1){10}}
\put(8,0){$\bullet$}\put(28,0){$\bullet$}
\put(44,0){=}
\put(58,0){$\bullet$}
\put(66,0){\line(0,1){10}\put(0,10){\vector(1,0){18}}\put(14,0){ \line(0,1){10}}}
\put(86,0){$\bullet$}
\put(96,0){+}
\put(110,0){\line(0,1){10}\put(0,10){\vector(1,0){20}}\put(16,0){ \line(0,1){10}}}
\put(134,0){$\bullet$}
\put(154,0){$\bullet$}
}
 \\
\put(-100,0){
\put(8,0){\line(0,1){10}\put(0,10){\vector(1,0){20}}\put(16,0){ \line(0,1){10}}}
$\bullet$\put(10,0){$\bullet$}
\put(44,0){=}
\put(58,0){$\bullet$}
\put(66,0){\line(0,1){10}\put(0,10){\vector(1,0){18}}\put(14,0){ \line(0,1){10}}}
\put(86,0){$\bullet$}
\put(96,0){+}
\put(134,0){\line(0,1){10}\put(0,10){\vector(1,0){20}}\put(16,0){ \line(0,1){10}}}
\put(110,0){$\bullet$}
\put(124,0){$\bullet$}
}
\nonumber \\
\put(-100,0){
\line(0,1){10}\put(0,10){\vector(1,0){28}}\put(24,0){ \line(0,1){10}}
\put(4,0){$\bullet$}\put(18,0){$\bullet$}
\put(44,0){=}
\put(58,0){$\bullet$}
\put(66,0){\line(0,1){10}\put(0,10){\vector(1,0){18}}\put(14,0){ \line(0,1){10}}}
\put(86,0){$\bullet$}
\put(96,0){+}
\put(134,0){\line(0,1){10}\put(0,10){\vector(1,0){20}}\put(16,0){ \line(0,1){10}}}
\put(110,0){$\bullet$}
\put(124,0){$\bullet$}
\put(158,0){+}
\put(170,0){\line(0,1){10}\put(0,10){\vector(1,0){20}}\put(16,0){ \line(0,1){10}}}
\put(192,0){$\bullet$}
\put(204,0){$\bullet$}
\put(215,0){.}} \nonumber
\end{eqnarray}

\section{Scalar product}\label{product}
First we note that only states with equal total spin, i.e.
states with equal number of singlet $S$, triplet $T$ and quintet $Q$ bonds, have a
non-vanishing scalar product.  Using Eq.~(\ref{notation}) 
we write the hws (\ref{hws})  in the expanded form
\begin{eqnarray}\label{asp}
& & \ket{\psi_s}  =  \frac{1}{\sqrt{2}^T\sqrt{3}^S}\sum_{(t_1,t_2)\in \tau, 
(s_1,s_2)\in \Sigma}\ket{+^{(i_1)}..+^{(i_{2Q})}}  \\
 & & \times 
\ket{t_1^{i_{(2Q+1)}}t_2^{i_{(2Q+2)}}}\ket{s_1^{(i_{2Q+3})}s_2^{(i_{2Q+4})}}..\ket{s_1^{(i_{N-1})}s_2^{(i_{N})}}\ , \nonumber
\end{eqnarray}
where $\tau \equiv \{(+,0),-(0,+)\}$ and $\Sigma\equiv \{(+,-),(-,+),-(0,0)\}$. 
The algorithm how to compute the scalar product is 
based on determining the number of different
combinations of $S^z$ product states in the decomposition (\ref{asp}) 
that are the same in the two states in the
scalar product.  
This number multiplied by the corresponding 
normalization factors gives the scalar product.

For illustration let us consider 
two states in a chain of  length $N=10$
as an example 
\begin{eqnarray}
\ket{1}&=&(1,2,3,4)[5,6]\{7,8\}\{9,10\} \nonumber \\ 
\ket{2}&=&(1,2,5,7)[3,6]\{4,8\}\{9,10\}\ . \nonumber 
\end{eqnarray} 
In order to calculate the scalar product $\langle 1|2\rangle$ we simply count the 
number of states that agree on both sides after the decomposition (\ref{asp}) and 
then multiply by the normalization.
The quintet spins $(1, 2, 3, 4)$ on the left hand side and $(1, 2,5,7)$ 
on the right hand side must all be in the state $|+\rangle$.
Therefore, according to Eq.~(\ref{notation}) only states can agree on both sides
of the scalar product where the $S^z$ 
eigenvalues of the spins at site $6$  is $|0\rangle$ and at site $8$ it
must be $|-\rangle$.  
We have exactly one possibility of different combinations of product states
to agree, which must be multiplied by the respective normalizations 
in Eq.~(\ref{notation}).
The spins with the label $9$ and $10$ are 
already in the same state and do not change the scalar product (or alternatively speaking 
this bond contributes a factor of 3 both in the numerator and denominator).
Therefore, we obtain after multiplying by the normalization factors
$\langle 1|2\rangle =3/(2\cdot3^2)=1/(2\cdot3)$. 
Obviously, scalar products 
factorize according to ''connected clusters'' and 
the analysis can be done for each connected cluster separately.  From 
this it also follows that in order for the scalar product to be non-zero,
at least one spin of each singlet bond 
must again be in a singlet bond on the other side of the scalar product.

The application of this algorithm on connected clusters with $s=0$
gives a scalar product which is determined by
the minimum number of index exchanges $\gamma$ between bonds in order to 
bring the two states into the same hws state
\begin{eqnarray}
& & \langle \psi_{s=0}(i_1,...,i_N)|\psi_{s=0}(i_1',...,i_N')\rangle =  \\
& & \ \ \ \ \ \ \ 
\langle\psi_{s=0}(i_1,...,i_N) | \overbrace{V...V}^{\gamma}|\psi_{s=0}(i_1,...,i_N)\rangle 
=\frac{1}{3^{\gamma}}\ , \nonumber
\end{eqnarray}
where $V_{i,j}\{.,i\}\{j,.\}=\{.,j\}\{i,.\}$ 
exchanges the indices between two singlet bonds.
This rule can most easily be understood by first considering two equal states with
$\langle \psi|\psi\rangle=1$.  Then by 
successively changing indices with $V_{i,j}$ in order to create
$|\psi'\rangle = \overbrace{V...V}^{\gamma}|\psi\rangle$ it is clear that the 
product state with $|00\rangle$ 
remains the same, while product states involving $|+-\rangle$ no longer agree.
Each application of $V$ therefore reduces the 
scalar product by a factor of three.
In a $s=0$ connected cluster of $N$ sites, there are $N/2$ singlet bonds.  The exchange 
operator $V$ can at most be applied $N/2-1$ times between bonds 
in order to bring the states into an equivalent form.  Therefore, the scalar product must 
be at least $1/3^{N/2-1}$ or larger between such states.

As an example let us consider the states
\begin{eqnarray} 
\ket{3}&=&\{1,2\}\{3,4\}\{5,6\}\{7,8\}\{9,10\}\nonumber \\
\ket{4}&=&\{1,10\}\{2,5\}\{4,6\}\{7,8\}\{3,9\}\ .\nonumber 
\end{eqnarray} 
The scalar product 
$\langle 3|4\rangle$ can be obtained from the number of exchange operators 
which we need to transform the state 
$\ket{4}$ into the state $\ket{3}$. 
\begin{equation}
\langle 3|4\rangle=\langle 3|V_{3,6}V_{5,9}V_{2,10}|3\rangle =\frac{1}{3^3}\ .
\end{equation}
There are other possible choices of exchange operators $V_{i,j}$, 
but the minimum number of exchanges is
always $\gamma=3$ in this case.

\nopagebreak


%

\end{document}